\documentclass[twoside,12pt]{article}
\usepackage{indentfirst}
\usepackage{bm}
\usepackage{amsmath}
\usepackage{CJK}
\usepackage{mathrsfs}
\usepackage{amsfonts}

\begin{document}

\begin{center}
\LARGE\bf New Relativistic Wave Equations for Two-Particle Systems
\end{center}

\footnotetext{\hspace*{-.45cm}\footnotesize $^\dag$ Corresponding
author. E-mail: yananbiguangqing@163.com
\\${}^{\rm b)}$ E-mail: yuekaifly@163.com}

\begin{center}
\rm Guang-Qing Bi $^{\rm a)\dagger}$, \ \ Yue-Kai Bi $^{\rm b)}$
\end{center}

\vspace*{2mm}

\begin{center}
\begin{minipage}{13.5cm}
\parindent 20pt\footnotesize
We seek to introduce a mathematical method to derive the
relativistic wave equations for two-particle system. According to
this method, if we define stationary wave functions as special
solutions like
$\Psi(\mathbf{r}_1,\mathbf{r}_2,t)=\psi(\mathbf{r}_1,\mathbf{r}_2)e^{-iEt/\hbar},\, \psi(\mathbf{r}_1,\mathbf{r}_2)\in\mathscr{S}(\mathbb{R}^3\times\mathbb{R}^3)$,
and properly define the relativistic reduced mass $\mu_0$, then some
new relativistic two-body wave equations can be derived. On this basis, we obtain the two-body Sommerfeld fine-structure formula for relativistic atomic two-body systems such as the pionium and pionic hydrogen atoms bound states, using which, we discuss the pair production
and annihilation of $\pi+$ and $\pi-$.
\end{minipage}
\end{center}

\begin{center}
\begin{minipage}{15.5cm}
\begin{minipage}[t]{2.3cm}{\bf Keywords:}\end{minipage}
\begin{minipage}[t]{13.1cm}
Relativistic two-body wave equations, Two-body Sommerfeld fine-structure formula
\end{minipage}\par\vglue8pt
{\bf PACS: } 03.65.Pm, 03.65.Ge
\end{minipage}
\end{center}

\section{Introduction}

The lack of an analytically solvable relativistic wave
equation for two-body atomic systems has compelled
physicists to use second-order perturbation theory in calculating
energy levels to order $\alpha^6$ in systems such as
positronium (See \cite{za08}, \cite{cz99}). In Reference \cite{co91} and \cite{co12} a two-particle Sommerfeld fine-structure formula was derived from the Bethe-Salpeter equation for two spin-$1/2$ constituent particles bound by a single-photon-exchange kernel in the Coulomb gauge. It is
\begin{equation}\label{yco}
E=\sqrt{m^2+M^2+\frac{2mM}{\sqrt{1+\frac{Z^2\alpha^2}{(n+\epsilon+1)^2}}}}\,c^2.
\end{equation}
Here $\alpha$ is the fine-structure constant, $n$ is the radial quantum number, $m$ is the mass of the electron and $M$ is the mass of the other particle.
In this work we will not be able to derive an angular
equation for $\epsilon$. Nevertheless, from the two-particle Sommerfeld
formula (\ref{yco}) alone, without knowing $\epsilon$, we have found
two predictions at order $\alpha^6$ which are verified in particular
atomic systems by previous calculations which used
second-order perturbation theory. These results suggest
that it may soon be possible to find an analytically solvable
relativistic atomic two-body wave equation which
would eliminate the need for second-order perturbation
calculations to obtain energies to order $\alpha^6$.
Both special relativity and experiments indicate that, the mass of a
many-particle system in a bound state is less than the sum of the
rest mass of every particle forming the system, and the difference
gives the mass defect of the system, while the product of the mass
defect and the square of the speed of light gives the binding energy
of the system. As the binding energy is quantized, the sum of it and
the rest mass of every particle forming the system is the energy
level of the system. For instance, the mass of an atomic nucleus is
obviously less than the sum of the rest mass of every nucleon
forming the atomic nucleus. Therefore, in order to express the mass
defect explicitly, there is a necessity to introduce the concept of
system mass, which differs from the sum of the rest mass of every
particle forming the system. By introducing the concept of the
system mass and applying proper mathematical skills, the
relativistic wave equations for two-particle system is derived. On
this basis, let us properly define the relativistic reduced mass to
further derive the new relativistic two-body wave equations.
The main results of this paper are expressed as

1. The relativistic two-body wave equations
\begin{eqnarray*}
 E'\psi &=& -\,\frac{2(m_{01}\mu+m_{02}\mu_0)}{(m_0+m)(\mu_0+\mu)}\frac{\hbar^2}{m_{01}}\nabla_1^2\psi
 -\frac{2(m_{02}\mu+m_{01}\mu_0)}{(m_0+m)(\mu_0+\mu)}\frac{\hbar^2}{m_{02}}\nabla_2^2\psi\nonumber\\
 & &+\,\frac{2\mu}{\mu_0+\mu}\left(\frac{2m}{m_0+m}U-\frac{U^2}{(m_0+m)c^2}\right)\psi\nonumber\\
 & &-\,\frac{1}{(\mu_0+\mu)c^2}\left(\frac{2m}{m_0+m}U-\frac{U^2}{(m_0+m)c^2}\right)^2\psi\nonumber\\
 & &+\,\frac{\hbar^2}{(m_0+m)(\mu_0+\mu)c^2}\left(\frac{2m}{m_0+m}U-\frac{U^2}{(m_0+m)c^2}\right)(\nabla_1^2+\nabla_2^2)\psi\nonumber\\
 & &+\,\frac{\hbar^2}{(m_0+m)(\mu_0+\mu)c^2}(\nabla_1^2+\nabla_2^2)\left[\left(\frac{2m}{m_0+m}U-\frac{U^2}{(m_0+m)c^2}\right)\psi\right]\nonumber\\
 & &-\,\frac{\hbar^4}{(m_0+m)^2(\mu_0+\mu)c^2}(\nabla_1^2-\nabla_2^2)^2\psi.
\end{eqnarray*}

2. In the center-of-momentum frame, it is simplified as
\begin{eqnarray*}
 E'\psi
 &=&-\,\frac{4\hbar^2}{(m_0+m)^2(\mu_0+\mu)}\left(m-\frac{U}{c^2}\right)^2\nabla^2\psi
 +\frac{2\mu}{(\mu_0+\mu)(m_0+m)}\left(2mU-\frac{U^2}{c^2}\right)\psi\nonumber\\
 & &+\,\frac{2\hbar^2}{(m_0+m)^2(\mu_0+\mu)c^2}\left[\nabla^2\left(2mU-\frac{U^2}{c^2}\right)\right]\psi\nonumber\\
 & &+\,\frac{4\hbar^2}{(m_0+m)^2(\mu_0+\mu)c^2}\,\nabla\left(2mU-\frac{U^2}{c^2}\right)\cdot\nabla\psi\nonumber\\
 & &-\,\frac{1}{(m_0+m)^2(\mu_0+\mu)c^2}\left(2mU-\frac{U^2}{c^2}\right)^2\psi.
\end{eqnarray*}
Where $m_0=m_{01}+m_{02},\,E=mc^2$, and $m,\mu_0,\mu$ respectively
denote
\[m=m_0+\frac{1}{c^2}E',\quad\mu_0=\frac{2m_{01}m_{02}}{m_0+m},\quad\mu=\mu_0+\frac{1}{c^2}E'.\]

3. If
$U=-Ze_s^2/r,\;e_s=e(4\pi\varepsilon_0)^{-1/2},\;r=|\mathbf{r}_1-\mathbf{r}_2|$
and $E'<0$, then
\begin{equation*}
   E_n=\left[m_{01}^2\pm2m_{01}m_{02}\left(1+\frac{Z^2\alpha^2}{(n-\sigma_l)^2}\right)^{-1/2}+m_{02}^2\right]^{1/2}c^2.
\end{equation*}
\begin{equation*}
    \sigma_l=l+\frac{1}{2}+\frac{d_0}{2(n-\sigma_l)}-\sqrt{\left(l+\frac{1}{2}\right)^2-Z^2\alpha^2+\frac{3d_0}{2}-\frac{d_0}{n-\sigma_l}},
    \quad\alpha=\frac{e_s^2}{\hbar c}.
\end{equation*}
\begin{equation*}
   d_0=2Z^2\alpha^2D,\quad{D}=\frac{\mu(m_0+m)}{2m^2},\quad m=E_n/c^2,\quad\mu=\pm\mu_0\left(1+\frac{Z^2\alpha^2}{(n-\sigma_l)^2}\right)^{-1/2}.
\end{equation*}

\section{Relativistic Two-body Wave Equations}

As we know, arbitrary wave function is equal to the linear
superposition of the plane waves of free particles with all possible
momentum, namely

Let $E$ be the total energy of the system, $\mathbf{p}$ be the momentum of particle, then

\parbox{14cm}{\begin{eqnarray*}\label{1a}
 \Psi(\mathbf{r},t)&=&\int\!\!\!\int\limits^\infty_{-\infty}\!\!\!\int{c}(\mathbf{p},t)\Psi_p(\mathbf{r},t)\,dp_xdp_ydp_z,\\
c(\mathbf{p},t)&=&\int\!\!\!\int\limits^\infty_{-\infty}\!\!\!\int\Psi(\mathbf{r},t)\Psi_p^*(\mathbf{r},t)\,dx\,dy\,dz.
\end{eqnarray*}}\hfill\parbox{1cm}{\begin{eqnarray}\end{eqnarray}}
Where $\Psi_p$ is
\begin{equation}\label{1}
\Psi_p(\mathbf{r},t)=A\exp(-i(Et-\mathbf{p\cdot{r}})/\hbar).
\end{equation}
For the one-particle system, $A=(2\pi\hbar)^{-3/2}$.
Clearly, (\ref{1a}) are the Fourier transform and its inversion, which can be
extended to many-particle systems. For two-particle systems, let
$\mathbf{r}_1=(x_1,y_1,z_1)$ and $\mathbf{r}_2=(x_2,y_2,z_2)$ be
position vectors of two particles in the laboratory reference frame
respectively, and corresponding momentum vectors be
$\mathbf{p}_1=(p_{x_1},p_{y_1},p_{z_1})$ and
$\mathbf{p}_2=(p_{x_2},p_{y_2},p_{z_2})$. Thus related physical
quantities in (\ref{1a}) and (\ref{1}) are extended to
$\mathbf{r}=(\mathbf{r}_1,\mathbf{r}_2)$, $\mathbf{p}=(\mathbf{p}_1,\mathbf{p}_2)$, $A=(2\pi\hbar)^{-3}$, $dp_x=dp_{x_1}dp_{x_2}$,
$dp_y=dp_{y_1}dp_{y_2}$, $dp_z=dp_{z_1}dp_{z_2}$, $dx=dx_1dx_2,\,dy=dy_1dy_2,\,dz=dz_1dz_2$.

Assuming that any particle with the rest mass $m_0$, no matter how
high the speed is, no matter it is in a potential field or in free
space, and no matter how it interacts with other particles, its
kinetic energy is:
\begin{eqnarray}\label{2a}
E_k=(c^2p^2+m_0{}^2c^4)^{1/2}-m_0c^2.
\end{eqnarray}
On this basis, we can establish the relation between the system energy
$E$ and the momentum $\mathbf{p}$ using proper mathematical skills,
thus obtain the relativistic Hamiltonian. Therefore, we introduce
the mathematical method of Reference \cite{bi97}-\cite{bi11} to quantum mechanics. Similarly to Reference \cite{bi97}, we can introduce the relevant concepts in quantum mechanics:

\textbf{Definition 1} $\Psi_p(\mathbf{r},t)$ in the right-hand side of (\ref{1a}) is defined as
the base function of quantum mechanics, where $E$ and $\mathbf{p}$
are called the characters of base functions, while $E$ and
$\mathbf{p}$ are not only suitable for free particles, but also
suitable for any system, and the relation between $E$ and
$\mathbf{p}$ is called the characteristic equation of wave
equations. Different system has different characteristic equations.

According to differential laws, we have
\begin{equation}\label{2}
i\hbar\frac{\partial}{\partial{t}}\Psi_p=E\Psi_p,\quad{-i}\hbar\nabla_j\Psi_p=\mathbf{p}_j\Psi_p,\;j=1,2,\cdots.
\end{equation}

\textbf{Definition 2} Let $m_0=m_{01}+m_{02}+\cdots+m_{0N}$ be the
total rest mass of an $N$-particle system, $E'$ be the sum of the
kinetic energy and potential energy of all the $N$ particles, then
the actual mass of the system, which is called the system mass, is
defined as
\begin{equation}\label{3}
m=m_0+\frac{1}{c^2}E'.
\end{equation}

\textbf{Definition 3} If the system is in a bound state ($E'<0$),
then the absolute value of $E'$ is
\[|E'|=m_0c^2-mc^2=\triangle{m}c^2,\]
which is called the binding energy of the system, where
$\triangle{m}=m_0-m$ is the mass defect of the system.

\textbf{Definition 4} The total energy of the system $E$ is defined
as the sum of the rest energy, kinetic energy and potential energy
of all the particles forming the system, namely $E=m_0c^2+E'$го

According to Definition 2 and 4, the total energy of the system is
equal to the product of the system mass and the square of the speed
of light, namely $E=mc^2$, thus the system mass is uniquely
determined by the energy levels of the system.

\textbf{Definition 5} In relativistic quantum mechanics, the
stationary wave function for two-particle system is defined as the
following special solution:
\begin{equation}\label{4'}
\Psi(\mathbf{r}_1,\mathbf{r}_2,t)=\psi(\mathbf{r}_1,\mathbf{r}_2)\exp(-iEt/\hbar),\quad \psi(\mathbf{r}_1,\mathbf{r}_2)\in\mathscr{S}(\mathbb{R}^3\times\mathbb{R}^3).
\end{equation}
Where $E$ is the total energy of the two-particle system.

Applying (\ref{1a}) to Definition 5, we have

\parbox{13.5cm}{\begin{eqnarray*}\label{1ab5}
 \Psi(\mathbf{r}_1,\mathbf{r}_2,t)&=&\frac{1}{(2\pi\hbar)^3}\int\!\!\!\int\limits^\infty_{-\infty}\!\!\!\int
 {c}(\mathbf{p}_1,\mathbf{p}_2)\exp\left(\frac{i}{\hbar}(\mathbf{p}_1\cdot\mathbf{r}_1+\mathbf{p}_2\cdot\mathbf{r}_2-Et)\right)\,dp_xdp_ydp_z,\\
c(\mathbf{p}_1,\mathbf{p}_2)&=&\frac{1}{(2\pi\hbar)^3}\int\!\!\!\int\limits^\infty_{-\infty}\!\!\!\int\psi(\mathbf{r}_1,\mathbf{r}_2)
\exp\left(-\frac{i}{\hbar}(\mathbf{p}_1\cdot\mathbf{r}_1+\mathbf{p}_2\cdot\mathbf{r}_2)\right)\,dx\,dy\,dz.
              \end{eqnarray*}}\hfill\parbox{1cm}{\begin{eqnarray}\end{eqnarray}}
Where $\mathbf{r}_1=(x_1,y_1,z_1)$ and $\mathbf{r}_2=(x_2,y_2,z_2)$
are position vectors of two particles in the laboratory reference
frame respectively, and $dx=dx_1dx_2,\,dy=dy_1dy_2,\,dz=dz_1dz_2$.
Corresponding momentum vectors are respectively
$\mathbf{p}_1=(p_{x_1},p_{y_1},p_{z_1})$ and
$\mathbf{p}_2=(p_{x_2},p_{y_2},p_{z_2})$, and
$dp_x=dp_{x_1}dp_{x_2}$, $dp_y=dp_{y_1}dp_{y_2}$,
$dp_z=dp_{z_1}dp_{z_2}$. Because of $\psi(\mathbf{r}_1,\mathbf{r}_2)\in\mathscr{S}(\mathbb{R}^3\times\mathbb{R}^3)$ it satisfies natural boundary conditions: $\psi(\mathbf{r}_1,\mathbf{r}_2)\rightarrow0,\;\mathbf{r}\rightarrow\infty$.

In relativistic quantum mechanics, due to the relativistic effect
that mass varies with speed, the center of mass system is no longer
a proper description framework, instead, we use the
center-of-momentum frame, which is a coordinate system that the
total momentum equals zero. If $v_1,\,v_2$ respectively denote the
speed of particles in the two-particle system, then their momentum
respectively are
\[\mathbf{p}_1=m_1\mathbf{v}_1=\frac{m_{01}\mathbf{v}_1}{\sqrt{1-(v_1/c)^2}},\quad
\mathbf{p}_2=m_2\mathbf{v}_2=\frac{m_{02}\mathbf{v}_2}{\sqrt{1-(v_2/c)^2}},\]
and $\mathbf{p}_1=-\mathbf{p}_2$. If $v$ denotes the relative speed
between two particles, then we can properly define the relativistic
reduced mass $\mu$ to make the relative momentum
$\mathbf{p}=\mu\mathbf{v}$ satisfy
$|\mathbf{p}_1|=|\mathbf{p}_2|=|\mathbf{p}|$, namely
\begin{equation}\label{1ab}
    p_{x_1}^2+p_{y_1}^2+p_{z_1}^2=p_{x_2}^2+p_{y_2}^2+p_{z_2}^2=p_x^2+p_y^2+p_z^2.
\end{equation}
In other words, the reduced mass $\mu$ can be determined using
(\ref{1ab}) and the relativistic velocity addition formula. As it is
related to speed, in order to distinguish it from another type of
reduced mass, we call this one the speed-type reduced mass. For
instance, if two particles of a two-particle system are restricted
to movement along the same line, then its speed-type reduced mass is
defined as
\begin{equation}\label{1ab'}
\mu=\frac{m_1m_2}{m_1+m_2}\left(1+\frac{v_1v_2}{c^2}\right).
\end{equation}
Where
\[m_1=\frac{m_{01}}{\sqrt{1-(v_1/c)^2}},\quad{m}_2=\frac{m_{02}}{\sqrt{1-(v_2/c)^2}}.\]

Therefore, using the center-of-momentum frame in (\ref{1ab5}), $\mathbf{p}_2=-\mathbf{p}_1,\;|\mathbf{p}_1|=|\mathbf{p}|$.
Substituting them into (\ref{1ab5}), we have

\parbox{14cm}{\begin{eqnarray*}\label{1ab5'}
 \psi(\mathbf{r}_1,\mathbf{r}_2)&=&\frac{1}{(2\pi\hbar)^3}\int\!\!\!\int\limits^\infty_{-\infty}\!\!\!\int
 {c}(\mathbf{p}_1,\mathbf{p}_2)\exp(i\mathbf{p}\cdot(\mathbf{r}_1-\mathbf{r}_2)/\hbar)\,dp_xdp_ydp_z,\\
c(\mathbf{p}_1,\mathbf{p}_2)&=&\frac{1}{(2\pi\hbar)^3}\int\!\!\!\int\limits^\infty_{-\infty}\!\!\!\int\psi(\mathbf{r}_1,\mathbf{r}_2)
\exp(-i\mathbf{p}\cdot(\mathbf{r}_1-\mathbf{r}_2)/\hbar)\,dx\,dy\,dz.
              \end{eqnarray*}}\hfill\parbox{1cm}{\begin{eqnarray}\end{eqnarray}}
Where $|\mathbf{p}|$ is the relative momentum of the two-particle
system. Relative coordinate is denoted by
$\mathbf{r}=\mathbf{r}_1-\mathbf{r}_2$, then the result can be
expressed by a more symmetric form

\parbox{11cm}{\begin{eqnarray*}\label{1ab5''}
 \Psi(\mathbf{r}_1,\mathbf{r}_2,t)&=&\frac{1}{(2\pi\hbar)^3}\int\!\!\!\int\limits^\infty_{-\infty}\!\!\!\int
 {c}(\mathbf{p}_1,\mathbf{p}_2)\exp(-i(Et-\mathbf{p\cdot{r}})/\hbar)\,dp_xdp_ydp_z,\\
c(\mathbf{p}_1,\mathbf{p}_2)&=&\frac{1}{(2\pi\hbar)^3}\int\!\!\!\int\limits^\infty_{-\infty}\!\!\!\int\Psi(\mathbf{r}_1,\mathbf{r}_2,t)
\exp(i(Et-\mathbf{p\cdot{r}})/\hbar)\,dx\,dy\,dz.
              \end{eqnarray*}}\hfill\parbox{1cm}{\begin{eqnarray}\end{eqnarray}}
Where
$\Psi(\mathbf{r}_1,\mathbf{r}_2,t)=\psi(\mathbf{r}_1,\mathbf{r}_2)\exp(-iEt/\hbar)$
is the relativistic stationary wave functions for the two-particle
system. Therefore, $\mathbf{p}=\mu\mathbf{v}$, which is the relative
momentum of the two particle system in the center-of-momentum frame,
is definitely equivalent to the differential operator
$-i\hbar\nabla$ with respect to the relative coordinate
$\mathbf{r}=\mathbf{r}_1-\mathbf{r}_2$, (\ref{2}) becomes
\begin{equation}\label{2d}
i\hbar\frac{\partial}{\partial{t}}\Psi_p=E\Psi_p,\quad{-i}\hbar\nabla\Psi_p=\mathbf{p}\Psi_p.
\end{equation}

Considering an isolated two-particle system, if the interaction
energy between two particles is denoted by
$U(\mathbf{r}_1,\mathbf{r}_2)$, then according to Definition 2 and
(\ref{2a}), we have
\begin{equation}\label{4}
    E'-U=(c^2p_1^2+m_{01}^2c^4)^{1/2}-m_{01}c^2+(c^2p_2^2+m_{02}^2c^4)^{1/2}-m_{02}c^2.
\end{equation}
Where $m_{01},\,m_{02}$ are the rest mass of two particles
respectively, and corresponding momentum are
$p_1=|\mathbf{p}_1|,\,p_2=|\mathbf{p}_2|$. This is the
characteristic equation of relativistic wave equations for the
two-particle system, thus we obtain
\begin{equation}\label{sa}
    [\sqrt{c^2p_1^2+m_{01}^2c^4}+\sqrt{c^2p_2^2+m_{02}^2c^4}+U]\psi=E\psi.
\end{equation}
This is the spin-less Salpeter equation (See \cite{sr}, \cite{gi}),
it is an important relativistic two-body wave equation. In order to
make it easier to solve the corresponding relativistic wave
equation, the characteristic equation (\ref{4}) should be
transformed to remove the fractional power, then we have
\begin{eqnarray}\label{5}
(E'-U+m_{01}c^2+m_{02}c^2)^2 &=& c^2p_1^2+m_{01}^2c^4+c^2p_2^2+m_{02}^2c^4\nonumber\\
 & & +\,2(c^2p_1^2+\,m_{01}^2c^4)^{1/2}(c^2p_2^2+m_{02}^2c^4)^{1/2}.
\end{eqnarray}
Expanding the left-hand side of (\ref{5}) and applying (\ref{3}), we
have
\begin{eqnarray}\label{6}
 & & (m_0+m)E'-2mU+U^2/c^2\nonumber\\
 &=& p_1^2+p_2^2+2(p_1^2+m_{01}^2c^2)^{1/2}(p_2^2+m_{02}^2c^2)^{1/2}-2m_{01}m_{02}c^2.
\end{eqnarray}
Further, removing the radical sign, we have
\begin{eqnarray}\label{6'}
 & & [(m_0+m)E'+2m_{01}m_{02}c^2-p_1^2-p_2^2-2mU+U^2/c^2]^2\nonumber\\
 &=& 4(p_1^2+m_{01}^2c^2)(p_2^2+m_{02}^2c^2).
\end{eqnarray}

\textbf{Definition 6} In relativistic quantum mechanics, a type of
relativistic reduced mass $\mu_0$ of two-particle systems is defined
as
\begin{equation}\label{3'}
  \mu_0=\frac{2m_{01}m_{02}}{m_0+m},\quad\mu=\mu_0+\frac{1}{c^2}E'.
\end{equation}
Where $m_0=m_{01}+m_{02}$, $m$ is the system mass of the
two-particle system, $E'$ is the sum of kinetic energy and potential
energy of the two particles, $\mu$ is called the system mass
corresponding to $\mu_0$. Unless otherwise stated, the reduced mass
referred in our paper from now on is defined in this way, which
should not be confused with the speed-type reduced mass mentioned
previously.

According to (\ref{3'}), we have
\begin{equation}\label{j0'}
   \frac{m_{01}\mu+m_{02}\mu_0}{m_{01}}+\frac{m_{02}\mu+m_{01}\mu_0}{m_{02}}=\frac{2m^2}{m_0+m}.
\end{equation}
\begin{equation}\label{j0'1}
 (m_0+m)E'+2m_{01}m_{02}c^2=(m_0+m)\mu{c}^2.
\end{equation}
\begin{equation}\label{j0'2}
 ((m_0+m)E'+2m_{01}m_{02}c^2)^2=(m_0+m)^2(\mu_0+\mu)c^2E'+4m_{01}^2m_{02}^2c^4.
\end{equation}
By (\ref{j0'1}) and (\ref{j0'2}), the characteristic equation (\ref{6'}) becomes
\begin{eqnarray}\label{6'1}
 & & 4(p_1^2p_2^2+m_{02}^2c^2p_1^2+m_{01}^2c^2p_2^2)\nonumber\\
 &=&(m_0+m)^2(\mu_0+\mu)c^2E'-2(m_0+m)\mu{c}^2(p_1^2+p_2^2+2mU-U^2/c^2)\nonumber\\
 & &+\,(p_1^2+p_2^2+2mU-U^2/c^2)^2.
\end{eqnarray}

According to (\ref{6'1}), the relativistic Hamiltonian of two-particle systems can be expressed as
\begin{eqnarray}\label{6'2}
 H=E=E'+m_0c^2 &=& \frac{2(m_{01}\mu+m_{02}\mu_0)}{(m_0+m)(\mu_0+\mu)}\frac{p_1^2}{m_{01}}+\frac{2(m_{02}\mu+m_{01}\mu_0)}{(m_0+m)(\mu_0+\mu)}\frac{p_2^2}{m_{02}}\nonumber\\
 & &+\,\frac{2\mu}{\mu_0+\mu}\left(\frac{2m}{m_0+m}U-\frac{U^2}{(m_0+m)c^2}\right)\nonumber\\
 & &-\,\frac{1}{(m_0+m)(\mu_0+\mu)c^2}(p_1^2+p_2^2)\left(\frac{2m}{m_0+m}U-\frac{U^2}{(m_0+m)c^2}\right)\nonumber\\
 & &-\,\frac{1}{(m_0+m)(\mu_0+\mu)c^2}\left(\frac{2m}{m_0+m}U-\frac{U^2}{(m_0+m)c^2}\right)(p_1^2+p_2^2)\nonumber\\
 & &-\,\frac{1}{(\mu_0+\mu)c^2}\left(\frac{2m}{m_0+m}U-\frac{U^2}{(m_0+m)c^2}\right)^2\nonumber\\
 & &-\,\frac{1}{(m_0+m)^2(\mu_0+\mu)c^2}(p_1^2-p_2^2)^2+m_0c^2.
\end{eqnarray}
Therefore, taking (\ref{6'2}) as the characteristic equation,
multiplying both sides of the equation by the base function
$\Psi_p(\mathbf{r},t)$, and by using (\ref{2}), we have
\begin{eqnarray*}
 i\hbar\frac{\partial\Psi_p}{\partial{t}}
 &=& -\,\frac{2(m_{01}\mu+m_{02}\mu_0)}{(m_0+m)(\mu_0+\mu)}\frac{\hbar^2}{m_{01}}\nabla_1^2\Psi_p
 -\frac{2(m_{02}\mu+m_{01}\mu_0)}{(m_0+m)(\mu_0+\mu)}\frac{\hbar^2}{m_{02}}\nabla_2^2\Psi_p\\
 & &+\,\frac{2\mu}{\mu_0+\mu}\left(\frac{2m}{m_0+m}U-\frac{U^2}{(m_0+m)c^2}\right)\Psi_p\\
 & &+\,\frac{\hbar^2}{(m_0+m)(\mu_0+\mu)c^2}(\nabla_1^2+\nabla_2^2)\left[\left(\frac{2m}{m_0+m}U-\frac{U^2}{(m_0+m)c^2}\right)\Psi_p\right]\\
 & &+\,\frac{\hbar^2}{(m_0+m)(\mu_0+\mu)c^2}\left(\frac{2m}{m_0+m}U-\frac{U^2}{(m_0+m)c^2}\right)(\nabla_1^2+\nabla_2^2)\Psi_p\\
 & &-\,\frac{1}{(\mu_0+\mu)c^2}\left(\frac{2m}{m_0+m}U-\frac{U^2}{(m_0+m)c^2}\right)^2\Psi_p\\
 & &-\,\frac{\hbar^4}{(m_0+m)^2(\mu_0+\mu)c^2}(\nabla_1^2-\nabla_2^2)^2\Psi_p+m_0c^2\Psi_p.
\end{eqnarray*}
From (\ref{1ab5}) here $\Psi_p$ is expressed as
\begin{equation*}
 \Psi_p(\mathbf{r}_1,\mathbf{r}_2,t)=\frac{1}{(2\pi\hbar)^3}\exp\left(\frac{i}{\hbar}(\mathbf{p}_1\cdot\mathbf{r}_1+\mathbf{p}_2\cdot\mathbf{r}_2-Et)\right).
\end{equation*}
Where $U(\mathbf{r}_1,\mathbf{r}_2)$ denotes the potential energy of
the interaction between two particles, $\nabla_1^2,\,\nabla_2^2$ are
Laplace operators respectively corresponding to
$\mathbf{r}_1,\mathbf{r}_2$.
According to (\ref{1ab5}), in the operator equation which is tenable for
the base function $\Psi_p(\mathbf{r}_1,\mathbf{r}_2,t)$, as long as each operator
in the operator equation is a linear operator and each linear
operator does not explicitly contain the characters $E,\mathbf{p}_1$ and
$\mathbf{p}_2$ of $\Psi_p(\mathbf{r}_1,\mathbf{r}_2,t)$, then this operator equation
is also tenable for an arbitrary wave function $\Psi(\mathbf{r}_1,\mathbf{r}_2,t)$.
Whereas, considering that the system mass $m$ is equivalent to the
character $E$, this operator equation is not tenable for arbitrary
wave functions, but tenable for an stationary wave function like
(\ref{4'}). In other words, if $i\hbar\frac{\partial}{\partial{t}}\Psi_p=H\Psi_p$, from (\ref{1ab5}) we have
\begin{eqnarray*}
  H\Psi(\mathbf{r}_1,\mathbf{r}_2,t) &=& H\int\!\!\!\int\limits^\infty_{-\infty}\!\!\!\int
 {c}(\mathbf{p}_1,\mathbf{p}_2)\Psi_p\,dp_xdp_ydp_z=\int\!\!\!\int\limits^\infty_{-\infty}\!\!\!\int
 {c}(\mathbf{p}_1,\mathbf{p}_2)H\Psi_p\,dp_xdp_ydp_z\\
 &=& \int\!\!\!\int\limits^\infty_{-\infty}\!\!\!\int{c}(\mathbf{p}_1,\mathbf{p}_2)i\hbar\frac{\partial}{\partial{t}}\Psi_p\,dp_xdp_ydp_z
  =i\hbar\frac{\partial}{\partial{t}}\int\!\!\!\int\limits^\infty_{-\infty}\!\!\!\int{c}(\mathbf{p}_1,\mathbf{p}_2)\Psi_p\,dp_xdp_ydp_z\\
 &=& i\hbar\frac{\partial}{\partial{t}}\Psi(\mathbf{r}_1,\mathbf{r}_2,t),\quad \forall\psi(\mathbf{r}_1,\mathbf{r}_2)\in\mathscr{S}(\mathbb{R}^3\times\mathbb{R}^3).
\end{eqnarray*}
So we get the following results:

An isolated two-particle system, the total spin angular momentum of
which is zero, is described by the stationary wave function
$\Psi(\mathbf{r}_1,\mathbf{r}_2,t)$ or
$\psi(\mathbf{r}_1,\mathbf{r}_2)$, any stationary wave function
\begin{equation*}
\Psi(\mathbf{r}_1,\mathbf{r}_2,t)=\psi(\mathbf{r}_1,\mathbf{r}_2)\exp(-iEt/\hbar),\quad \psi(\mathbf{r}_1,\mathbf{r}_2)\in\mathscr{S}(\mathbb{R}^3\times\mathbb{R}^3)
\end{equation*}
satisfies the following relativistic wave equation:
\begin{eqnarray}\label{7}
 i\hbar\frac{\partial\Psi}{\partial{t}}
 &=& -\,\frac{2(m_{01}\mu+m_{02}\mu_0)}{(m_0+m)(\mu_0+\mu)}\frac{\hbar^2}{m_{01}}\nabla_1^2\Psi
 -\frac{2(m_{02}\mu+m_{01}\mu_0)}{(m_0+m)(\mu_0+\mu)}\frac{\hbar^2}{m_{02}}\nabla_2^2\Psi\nonumber\\
 & &+\,\frac{2\mu}{\mu_0+\mu}\left(\frac{2m}{m_0+m}U-\frac{U^2}{(m_0+m)c^2}\right)\Psi\nonumber\\
 & &-\,\frac{1}{(\mu_0+\mu)c^2}\left(\frac{2m}{m_0+m}U-\frac{U^2}{(m_0+m)c^2}\right)^2\Psi\nonumber\\
 & &+\,\frac{\hbar^2}{(m_0+m)(\mu_0+\mu)c^2}\left(\frac{2m}{m_0+m}U-\frac{U^2}{(m_0+m)c^2}\right)(\nabla_1^2+\nabla_2^2)\Psi\nonumber\\
 & &+\,\frac{\hbar^2}{(m_0+m)(\mu_0+\mu)c^2}(\nabla_1^2+\nabla_2^2)\left[\left(\frac{2m}{m_0+m}U-\frac{U^2}{(m_0+m)c^2}\right)\Psi\right]\nonumber\\
 & &-\,\frac{\hbar^4}{(m_0+m)^2(\mu_0+\mu)c^2}(\nabla_1^2-\nabla_2^2)^2\Psi+m_0c^2\Psi.
\end{eqnarray}
Where $m_0=m_{01}+m_{02},\,E'=E-m_0c^2,\,E=mc^2$. $m,\mu_0,\mu$ respectively denote
\[m=m_0+\frac{1}{c^2}E',\quad\mu_0=\frac{2m_{01}m_{02}}{m_0+m},\quad\mu=\mu_0+\frac{1}{c^2}E'.\]

Clearly, for non-relativistic limits, we have
\[\mu\rightarrow\mu_0\rightarrow\frac{m_{01}m_{02}}{m_0}=\frac{m_{01}m_{02}}{m_{01}+m_{02}}.\]

In other words, the relativistic wave function
$\psi(\mathbf{r}_1,\mathbf{r}_2)$ for two-particle systems is
determined by the following relativistic wave equation and natural
boundary conditions:
\begin{eqnarray}\label{9}
 E'\psi &=& -\,\frac{2(m_{01}\mu+m_{02}\mu_0)}{(m_0+m)(\mu_0+\mu)}\frac{\hbar^2}{m_{01}}\nabla_1^2\psi
 -\frac{2(m_{02}\mu+m_{01}\mu_0)}{(m_0+m)(\mu_0+\mu)}\frac{\hbar^2}{m_{02}}\nabla_2^2\psi\nonumber\\
 & &+\,\frac{2\mu}{\mu_0+\mu}\left(\frac{2m}{m_0+m}U-\frac{U^2}{(m_0+m)c^2}\right)\psi\nonumber\\
 & &-\,\frac{1}{(\mu_0+\mu)c^2}\left(\frac{2m}{m_0+m}U-\frac{U^2}{(m_0+m)c^2}\right)^2\psi\nonumber\\
 & &+\,\frac{\hbar^2}{(m_0+m)(\mu_0+\mu)c^2}\left(\frac{2m}{m_0+m}U-\frac{U^2}{(m_0+m)c^2}\right)(\nabla_1^2+\nabla_2^2)\psi\nonumber\\
 & &+\,\frac{\hbar^2}{(m_0+m)(\mu_0+\mu)c^2}(\nabla_1^2+\nabla_2^2)\left[\left(\frac{2m}{m_0+m}U-\frac{U^2}{(m_0+m)c^2}\right)\psi\right]\nonumber\\
 & &-\,\frac{\hbar^4}{(m_0+m)^2(\mu_0+\mu)c^2}(\nabla_1^2-\nabla_2^2)^2\psi.
\end{eqnarray}

For bound states, the total energy $E$ of the system is quantized,
which is called the system energy level. The system mass $m=E/c^2$
is uniquely determined by the system energy level $E$. Clearly, for
non-relativistic limits, this equation turns out to be the
Schr\"{o}dinger equation of two-particle systems. If the system is
in the external field, then the system potential energy
$U(\mathbf{r}_1,\mathbf{r}_2)$ includes both the potential energy of
the system in the external field and the interaction energy between
particles.

Using the center-of-momentum frame, then according to (\ref{1ab}) we have
$p_1^2=p_2^2=p^2$, where $p$ is the relative momentum. Considering (\ref{j0'}) or
\begin{equation}\label{j0}
   \frac{2(m_{01}\mu+m_{02}\mu_0)}{(m_0+m)(\mu_0+\mu)m_{01}}+\frac{2(m_{02}\mu+m_{01}\mu_0)}{(m_0+m)(\mu_0+\mu)m_{02}}=\frac{4m^2}{(\mu_0+\mu)(m_0+m)^2},
\end{equation}
then in the center-of-momentum frame, (\ref{6'2}) becomes
\begin{eqnarray}\label{10}
 H=E=E'+m_0c^2 &=& \left(\frac{2m}{m_0+m}\right)^2\frac{p^2}{\mu_0+\mu}+\frac{2\mu}{\mu_0+\mu}\left(\frac{2m}{m_0+m}U-\frac{U^2}{(m_0+m)c^2}\right)\nonumber\\
 & &-\,\frac{2}{(m_0+m)c^2}\frac{p^2}{\mu_0+\mu}\left(\frac{2m}{m_0+m}U-\frac{U^2}{(m_0+m)c^2}\right)\nonumber\\
 & &-\,\frac{2}{(m_0+m)c^2}\left(\frac{2m}{m_0+m}U-\frac{U^2}{(m_0+m)c^2}\right)\frac{p^2}{\mu_0+\mu}\nonumber\\
 & &-\,\frac{1}{(\mu_0+\mu)c^2}\left(\frac{2m}{m_0+m}U-\frac{U^2}{(m_0+m)c^2}\right)^2+m_0c^2.
\end{eqnarray}

Taking (\ref{10}) as the characteristic equation, similarly we have:
Considering an isolated two-particle system in the center-of-momentum frame, if the total spin angular momentum of the system is zero,
then the stationary wave function
$$\Psi(\mathbf{r},t)=\psi(\mathbf{r})\exp(-iEt/\hbar)$$
is determined by the following relativistic wave equation and natural boundary conditions:
\begin{eqnarray}\label{11}
 i\hbar\frac{\partial\Psi}{\partial{t}}
 &=& -\,\left(\frac{2m}{m_0+m}\right)^2\frac{\hbar^2}{\mu_0+\mu}\nabla^2\Psi
 +\frac{2\mu}{\mu_0+\mu}\left(\frac{2m}{m_0+m}U-\frac{U^2}{(m_0+m)c^2}\right)\Psi\nonumber\\
 & &+\,\frac{2}{(m_0+m)c^2}\frac{\hbar^2}{\mu_0+\mu}\nabla^2\left[\left(\frac{2m}{m_0+m}U-\frac{U^2}{(m_0+m)c^2}\right)\Psi\right]\nonumber\\
 & &+\,\frac{2}{(m_0+m)c^2}\left(\frac{2m}{m_0+m}U-\frac{U^2}{(m_0+m)c^2}\right)\frac{\hbar^2}{\mu_0+\mu}\nabla^2\Psi\nonumber\\
 & &-\,\frac{1}{(\mu_0+\mu)c^2}\left(\frac{2m}{m_0+m}U-\frac{U^2}{(m_0+m)c^2}\right)^2\Psi+m_0c^2\Psi.
\end{eqnarray}

\begin{eqnarray}\label{12}
 E'\psi &=& -\,\left(\frac{2m}{m_0+m}\right)^2\frac{\hbar^2}{\mu_0+\mu}\nabla^2\psi
 +\frac{2\mu}{\mu_0+\mu}\left(\frac{2m}{m_0+m}U-\frac{U^2}{(m_0+m)c^2}\right)\psi\nonumber\\
 & &+\,\frac{2}{(m_0+m)c^2}\frac{\hbar^2}{\mu_0+\mu}\nabla^2\left[\left(\frac{2m}{m_0+m}U-\frac{U^2}{(m_0+m)c^2}\right)\psi\right]\nonumber\\
 & &+\,\frac{2}{(m_0+m)c^2}\left(\frac{2m}{m_0+m}U-\frac{U^2}{(m_0+m)c^2}\right)\frac{\hbar^2}{\mu_0+\mu}\nabla^2\psi\nonumber\\
 & &-\,\frac{1}{(\mu_0+\mu)c^2}\left(\frac{2m}{m_0+m}U-\frac{U^2}{(m_0+m)c^2}\right)^2\psi.
\end{eqnarray}
These are also the expressions of relativistic wave equations (\ref{7}) and (\ref{9}) in the center-of-momentum frame respectively,
where $\nabla^2$ is the Laplace operator corresponding to the relative coordinate $\mathbf{r}=\mathbf{r}_1-\mathbf{r}_2$.

Relativistic wave equations (\ref{11}) and (\ref{12}) can be further expressed as
\begin{eqnarray}\label{13}
 i\hbar\frac{\partial\Psi}{\partial{t}}
 &=& -\,\left(\frac{2m}{m_0+m}\right)^2\frac{\hbar^2}{\mu_0+\mu}\nabla^2\Psi
 +\frac{2\mu}{(\mu_0+\mu)(m_0+m)}\left(2mU-\frac{U^2}{c^2}\right)\Psi\nonumber\\
 & &+\,\frac{2\hbar^2}{(m_0+m)^2(\mu_0+\mu)c^2}\left[\nabla^2\left(2mU-\frac{U^2}{c^2}\right)\right]\Psi\nonumber\\
 & &+\,\frac{4\hbar^2}{(m_0+m)^2(\mu_0+\mu)c^2}\nabla\left(2mU-\frac{U^2}{c^2}\right)\cdot\nabla\Psi\nonumber\\
 & &+\,\frac{4\hbar^2}{(m_0+m)^2(\mu_0+\mu)c^2}\left(2mU-\frac{U^2}{c^2}\right)\nabla^2\Psi\nonumber\\
 & &-\,\frac{1}{(m_0+m)^2(\mu_0+\mu)c^2}\left(2mU-\frac{U^2}{c^2}\right)^2\Psi+m_0c^2\Psi.
\end{eqnarray}

\begin{eqnarray}\label{14}
 E'\psi
 &=& -\,\left(\frac{2m}{m_0+m}\right)^2\frac{\hbar^2}{\mu_0+\mu}\nabla^2\psi
 +\frac{2\mu}{(\mu_0+\mu)(m_0+m)}\left(2mU-\frac{U^2}{c^2}\right)\psi\nonumber\\
 & &+\,\frac{2\hbar^2}{(m_0+m)^2(\mu_0+\mu)c^2}\left[\nabla^2\left(2mU-\frac{U^2}{c^2}\right)\right]\psi\nonumber\\
 & &+\,\frac{4\hbar^2}{(m_0+m)^2(\mu_0+\mu)c^2}\nabla\left(2mU-\frac{U^2}{c^2}\right)\cdot\nabla\psi\nonumber\\
 & &+\,\frac{4\hbar^2}{(m_0+m)^2(\mu_0+\mu)c^2}\left(2mU-\frac{U^2}{c^2}\right)\nabla^2\psi\nonumber\\
 & &-\,\frac{1}{(m_0+m)^2(\mu_0+\mu)c^2}\left(2mU-\frac{U^2}{c^2}\right)^2\psi.
\end{eqnarray}

On the right-hand side of (\ref{14}), if combining the first and the
fifth terms, the stationary relativistic wave equation for
two-particle systems in the center-of-momentum frame is expressed as
\begin{eqnarray}\label{16}
 E'\psi
 &=&-\,\frac{4\hbar^2}{(m_0+m)^2(\mu_0+\mu)}\left(m-\frac{U}{c^2}\right)^2\nabla^2\psi
 +\frac{2\mu}{(\mu_0+\mu)(m_0+m)}\left(2mU-\frac{U^2}{c^2}\right)\psi\nonumber\\
 & &+\,\frac{2\hbar^2}{(m_0+m)^2(\mu_0+\mu)c^2}\left[\nabla^2\left(2mU-\frac{U^2}{c^2}\right)\right]\psi\nonumber\\
 & &+\,\frac{4\hbar^2}{(m_0+m)^2(\mu_0+\mu)c^2}\,\nabla\left(2mU-\frac{U^2}{c^2}\right)\cdot\nabla\psi\nonumber\\
 & &-\,\frac{1}{(m_0+m)^2(\mu_0+\mu)c^2}\left(2mU-\frac{U^2}{c^2}\right)^2\psi.
\end{eqnarray}

As we know, the spin angular momentum not only has the general property of angular momentum $\mathbf{S}\times\mathbf{S}=i\hbar\mathbf{S}$, but also has its own particularity.
For instance, for electrons and protons, the projection of the spin angular momentum $\mathbf{S}$ in any direction only takes two values $\pm\hbar/2$.

For the convenience of studying this type of angular momentum, a type of dimensionless vector $\sigma$ is introduced, determined by
\begin{equation}\label{j2}
   \sigma\times\sigma=2i\sigma,\quad\sigma_x^2=\sigma_y^2=\sigma_z^2=1.
\end{equation}
Using this type of vector $\sigma$, the spin angular momentum is expressed as $\mathbf{S}=(\hbar/2)\sigma$.
According to (\ref{j2}), if $\mathbf{A},\,\mathbf{B}$ are two arbitrary vectors commuted with $\sigma$, we have
\begin{equation}\label{j4}
   (\sigma\cdot\mathbf{A})(\sigma\cdot\mathbf{B})=\mathbf{A}\cdot\mathbf{B}+i\sigma\cdot(\mathbf{A}\times\mathbf{B}).
\end{equation}

If the relativistic Hamiltonian of two-particle systems (\ref{6'2}) is rewritten as
\begin{eqnarray}\label{j7}
 E'&=& \frac{2(m_{01}\mu+m_{02}\mu_0)}{(m_0+m)(\mu_0+\mu)}\frac{p_1^2}{m_{01}}+\frac{2(m_{02}\mu+m_{01}\mu_0)}{(m_0+m)(\mu_0+\mu)}\frac{p_2^2}{m_{02}}\nonumber\\
 & &+\,\frac{2\mu}{\mu_0+\mu}\left(\frac{2m}{m_0+m}U-\frac{U^2}{(m_0+m)c^2}\right)\nonumber\\
 & &-\,\frac{2}{(m_0+m)^2(\mu_0+\mu)c^2}\left[\sigma_1\cdot\mathbf{p}_1\left(2mU-\frac{U^2}{c^2}\right)\right](\sigma_1\cdot\mathbf{p}_1)\nonumber\\
 & &-\,\frac{2}{(m_0+m)^2(\mu_0+\mu)c^2}\left[\sigma_2\cdot\mathbf{p}_2\left(2mU-\frac{U^2}{c^2}\right)\right](\sigma_2\cdot\mathbf{p}_2)\nonumber\\
 & &-\,\frac{2}{(m_0+m)^2(\mu_0+\mu)c^2}\left(2mU-\frac{U^2}{c^2}\right)(p_1^2+p_2^2)\nonumber\\
 & &-\,\frac{1}{(m_0+m)^2(\mu_0+\mu)c^2}\left[(p_1^2+p_2^2)\left(2mU-\frac{U^2}{c^2}\right)\right]\nonumber\\
 & &-\,\frac{1}{(\mu_0+\mu)c^2}\left(\frac{2m}{m_0+m}U-\frac{U^2}{(m_0+m)c^2}\right)^2\nonumber\\
 & &-\,\frac{1}{(m_0+m)^2(\mu_0+\mu)c^2}(p_1^2-p_2^2)^2.
\end{eqnarray}
Thus the stationary relativistic wave equation for two-particle systems can be expressed as

\begin{eqnarray}\label{j8}
 i\hbar\frac{\partial\Psi}{\partial{t}}
 &=& -\frac{2(m_{01}\mu+m_{02}\mu_0)}{(m_0+m)(\mu_0+\mu)}\frac{\hbar^2}{m_{01}}\nabla_1^2\Psi
 -\frac{2(m_{02}\mu+m_{01}\mu_0)}{(m_0+m)(\mu_0+\mu)}\frac{\hbar^2}{m_{02}}\nabla_2^2\Psi\nonumber\\
 & &+\,\frac{2\mu}{\mu_0+\mu}\left(\frac{2m}{m_0+m}U-\frac{U^2}{(m_0+m)c^2}\right)\Psi\nonumber\\
 &&+\,\frac{2\hbar^2}{(m_0+m)^2(\mu_0+\mu)c^2}\left[\sigma_1\cdot\nabla_1\left(2mU-\frac{U^2}{c^2}\right)\right](\sigma_1\cdot\nabla_1)\Psi\nonumber\\
 &&+\,\frac{2\hbar^2}{(m_0+m)^2(\mu_0+\mu)c^2}\left[\sigma_2\cdot\nabla_2\left(2mU-\frac{U^2}{c^2}\right)\right](\sigma_2\cdot\nabla_2)\Psi\nonumber\\
 & &+\,\frac{2\hbar^2}{(m_0+m)^2(\mu_0+\mu)c^2}\left(2mU-\frac{U^2}{c^2}\right)(\nabla_1^2+\nabla_2^2)\Psi\nonumber\\
 & &+\,\frac{\hbar^2}{(m_0+m)^2(\mu_0+\mu)c^2}\left[(\nabla_1^2+\nabla_2^2)\left(2mU-\frac{U^2}{c^2}\right)\right]\Psi\nonumber\\
 & &-\,\frac{1}{(\mu_0+\mu)c^2}\left(\frac{2m}{m_0+m}U-\frac{U^2}{(m_0+m)c^2}\right)^2\Psi\nonumber\\
 & &-\,\frac{\hbar^4}{(m_0+m)^2(\mu_0+\mu)c^2}(\nabla_1^2-\nabla_2^2)^2\Psi+m_0c^2\Psi.
\end{eqnarray}
The two particles of the system are spin $1/2$ particles, where $U(\mathbf{r}_1,\mathbf{r}_2)$ denotes the potential energy of the system in the external field and the interaction energy between particles.

Let $\mathbf{S}_1$ be the spin angular momentum of the first particle, and $\mathbf{S}_2$ be that of the second one. In the central field, according to (\ref{j4}), (\ref{j8}) can be express as

\begin{eqnarray}\label{j9}
 i\hbar\frac{\partial\Psi}{\partial{t}}
 &=& -\frac{2(m_{01}\mu+m_{02}\mu_0)}{(m_0+m)(\mu_0+\mu)}\frac{\hbar^2}{m_{01}}\nabla_1^2\Psi
 -\frac{2(m_{02}\mu+m_{01}\mu_0)}{(m_0+m)(\mu_0+\mu)}\frac{\hbar^2}{m_{02}}\nabla_2^2\Psi\nonumber\\
 & &+\,\frac{2\mu}{\mu_0+\mu}\left(\frac{2m}{m_0+m}\,U-\frac{U^2}{(m_0+m)c^2}\right)\Psi\nonumber\\
 &&+\,\frac{4\hbar^2(m-U/c^2)}{(m_0+m)^2(\mu_0+\mu)c^2}
 \left(\frac{d{U}}{dr_1}\frac{\partial\Psi}{\partial{r_1}}+\frac{d{U}}{dr_2}\frac{\partial\Psi}{\partial{r_2}}\right)\nonumber\\
 & &+\,\frac{2\hbar^2}{(m_0+m)^2(\mu_0+\mu)c^2}\left(2mU-\frac{U^2}{c^2}\right)(\nabla_1^2+\nabla_2^2)\Psi\nonumber\\
 & &+\,\frac{\hbar^2}{(m_0+m)^2(\mu_0+\mu)c^2}\left[(\nabla_1^2+\nabla_2^2)\left(2mU-\frac{U^2}{c^2}\right)\right]\Psi\nonumber\\
 & &-\,\frac{1}{(\mu_0+\mu)c^2}\left(\frac{2m}{m_0+m}\,U-\frac{U^2}{(m_0+m)c^2}\right)^2\Psi\nonumber\\
 &&-\,\frac{8(m-U/c^2)}{(m_0+m)^2(\mu_0+\mu)c^2}
 \left(\frac{1}{r_1}\frac{d{U}}{dr_1}\mathbf{S}_1\cdot\mathbf{L}_1\Psi+\frac{1}{r_2}\frac{d{U}}{dr_2}\mathbf{S}_2\cdot\mathbf{L}_2\Psi\right)\nonumber\\
 & &-\,\frac{\hbar^4}{(m_0+m)^2(\mu_0+\mu)c^2}\,(\nabla_1^2-\nabla_2^2)^2\Psi+m_0c^2\Psi.
\end{eqnarray}

\begin{eqnarray}\label{j10}
  E'\psi
 &=& -\frac{2(m_{01}\mu+m_{02}\mu_0)}{(m_0+m)(\mu_0+\mu)}\frac{\hbar^2}{m_{01}}\nabla_1^2\psi
 -\frac{2(m_{02}\mu+m_{01}\mu_0)}{(m_0+m)(\mu_0+\mu)}\frac{\hbar^2}{m_{02}}\nabla_2^2\psi\nonumber\\
 & &+\,\frac{2\mu}{\mu_0+\mu}\left(\frac{2m}{m_0+m}\,U-\frac{U^2}{(m_0+m)c^2}\right)\psi\nonumber\\
 & &+\,\frac{4\hbar^2(m-U/c^2)}{(m_0+m)^2(\mu_0+\mu)c^2}
 \left(\frac{d{U}}{dr_1}\frac{\partial\psi}{\partial{r_1}}+\frac{d{U}}{dr_2}\frac{\partial\psi}{\partial{r_2}}\right)\nonumber\\
 & &+\,\frac{2\hbar^2}{(m_0+m)^2(\mu_0+\mu)c^2}\left(2mU-\frac{U^2}{c^2}\right)(\nabla_1^2+\nabla_2^2)\psi\nonumber\\
 & &+\,\frac{\hbar^2}{(m_0+m)^2(\mu_0+\mu)c^2}\left[(\nabla_1^2+\nabla_2^2)\left(2mU-\frac{U^2}{c^2}\right)\right]\psi\nonumber\\
 & &-\,\frac{1}{(\mu_0+\mu)c^2}\left(\frac{2m}{m_0+m}\,U-\frac{U^2}{(m_0+m)c^2}\right)^2\psi\nonumber\\
 & &-\,\frac{8(m-U/c^2)}{(m_0+m)^2(\mu_0+\mu)c^2}
 \left(\frac{1}{r_1}\frac{d{U}}{dr_1}\mathbf{S}_1\cdot\mathbf{L}_1\psi+\frac{1}{r_2}\frac{d{U}}{dr_2}\mathbf{S}_2\cdot\mathbf{L}_2\psi\right)\nonumber\\
 & &-\,\frac{\hbar^4}{(m_0+m)^2(\mu_0+\mu)c^2}\,(\nabla_1^2-\nabla_2^2)^2\psi.
\end{eqnarray}
Where $\mathbf{L}_1$ is the orbital angular momentum of the first particle, and $\mathbf{L}_2$ is that of the second one.
$\Psi$ is the stationary relativistic wave function
\[\Psi(\mathbf{r}_1,\mathbf{r}_2,s_{1z},s_{2z},t)=\psi(\mathbf{r}_1,\mathbf{r}_2,s_{1z},s_{2z})\exp(-iEt/\hbar).\]

For an isolated two-particle system, in the center-of-momentum frame, $\mathbf{p}_2=-\mathbf{p}_1,\;|\mathbf{p}_1|=|\mathbf{p}|$,
and $|\mathbf{p}|$ is the relative momentum of the two-particle system.
Based on the corresponding relation between momentum operators and gradient operators, according to $\mathbf{p}_2=-\mathbf{p}_1=-\mathbf{p}$
we have $\nabla_2=-\nabla_1=-\nabla$. Where $\nabla_1,\;\nabla_2$ and $\nabla$ are gradient operators corresponding to the coordinates
$\mathbf{r}_1,\;\mathbf{r}_2$ and $\mathbf{r}=\mathbf{r}_1-\mathbf{r}_2$.

Therefore, if using the center-of-momentum frame in the wave equation (\ref{j9}), supposing an isolated two-particle system, then
Using $\nabla_2=-\nabla_1=-\nabla$ and $\mathbf{p}_2=-\mathbf{p}_1=-\mathbf{p}$, in the central field we have
\[\frac{d{U}}{dr_1}\frac{\partial\Psi}{\partial{r}_1}=(\nabla_1U)\cdot\nabla_1\Psi=(\nabla{U})\cdot\nabla\Psi
=\frac{d{U}}{dr}\frac{\partial\Psi}{\partial{r}}.\]
\[\frac{d{U}}{dr_2}\frac{\partial\Psi}{\partial{r}_2}=(\nabla_2U)\cdot\nabla_2\Psi=(-\nabla{U})\cdot(-\nabla\Psi)
=\frac{d{U}}{dr}\frac{\partial\Psi}{\partial{r}}.\]
\[\frac{1}{r_1}\frac{d{U}}{dr_1}\mathbf{L}_1=\frac{1}{r_1}\frac{d{U}}{dr_1}\mathbf{r}_1\times\mathbf{p}_1=(\nabla_1U)\times\mathbf{p}_1
=(\nabla{U})\times\mathbf{p}=\frac{1}{r}\frac{d{U}}{dr}\mathbf{r}\times\mathbf{p}.\]
\[\frac{1}{r_2}\frac{d{U}}{dr_2}\mathbf{L}_2=\frac{1}{r_2}\frac{d{U}}{dr_2}\mathbf{r}_2\times\mathbf{p}_2=(\nabla_2U)\times\mathbf{p}_2
=(-\nabla{U})\times(-\mathbf{p})=\frac{1}{r}\frac{d{U}}{dr}\mathbf{r}\times\mathbf{p}.\]
Considering (\ref{j0}), the wave equation (\ref{j9}) can be expressed as

\begin{eqnarray}\label{j12}
 i\hbar\frac{\partial\Psi}{\partial{t}}
 &=& -\frac{4m^2\hbar^2}{(\mu_0+\mu)(m_0+m)^2}\nabla^2\Psi
 +\frac{2\mu}{\mu_0+\mu}\left(\frac{2m}{m_0+m}\,U-\frac{U^2}{(m_0+m)c^2}\right)\Psi\nonumber\\
 & &+\,\frac{8\hbar^2(m-U/c^2)}{(m_0+m)^2(\mu_0+\mu)c^2}\frac{d{U}}{dr}\frac{\partial\Psi}{\partial{r}}\nonumber\\
 & &+\,\frac{4\hbar^2}{(m_0+m)^2(\mu_0+\mu)c^2}\left(2mU-\frac{U^2}{c^2}\right)\nabla^2\Psi\nonumber\\
 & &+\,\frac{2\hbar^2}{(m_0+m)^2(\mu_0+\mu)c^2}\left[\nabla^2\left(2mU-\frac{U^2}{c^2}\right)\right]\Psi\nonumber\\
 & &-\,\frac{1}{(\mu_0+\mu)c^2}\left(\frac{2m}{m_0+m}\,U-\frac{U^2}{(m_0+m)c^2}\right)^2\Psi\nonumber\\
 &&-\,\frac{8(m-U/c^2)}{(m_0+m)^2(\mu_0+\mu)c^2}\frac{1}{r}\frac{d{U}}{dr}(\mathbf{S}_1+\mathbf{S}_2)\cdot(\mathbf{r}\times\mathbf{p})\Psi+m_0c^2\Psi.
\end{eqnarray}
The total spin angular momentum of the system is $\mathbf{S}=\mathbf{S}_1+\mathbf{S}_2$, and the orbital angular momentum $\mathbf{L}$ is
\[\mathbf{L}=\mathbf{L}_1+\mathbf{L}_2=\mathbf{r}_1\times\mathbf{p}_1+\mathbf{r}_2\times\mathbf{p}_2
=(\mathbf{r}_1-\mathbf{r}_2)\times\mathbf{p}_1=\mathbf{r}\times\mathbf{p}.\]
Therefore, the total orbital angular momentum $\mathbf{L}$ of the two-particle system in the center-of-momentum frame
is equal to the cross product of the relative coordinate $\mathbf{r}$ and the relative momentum $\mathbf{p}$.
Combining the first and the fourth terms on the right-hand side of (\ref{j12}), we have

Let $\mathbf{S},\;\mathbf{L}$ be the total spin angular momentum and total orbital angular momentum operators of the two-particle system respectively,
$U(\mathbf{r})$ be the interaction energy between particles, then in the central field,
the stationary wave function for the two-particle system in the center-of-momentum frame \[\Psi(\mathbf{r},s_z,t)=\psi(\mathbf{r},s_z)\exp(-iEt/\hbar)\]
is determined by the following relativistic wave function and natural boundary conditions, namely

\begin{eqnarray}\label{j13}
 i\hbar\frac{\partial\Psi}{\partial{t}}
 &=& -\frac{4\hbar^2}{(\mu_0+\mu)(m_0+m)^2}\left(m-\frac{U}{c^2}\right)^2\nabla^2\Psi\nonumber\\
 & &+\,\frac{2\mu}{(\mu_0+\mu)(m_0+m)}\left(2mU-\frac{U^2}{c^2}\right)\Psi\nonumber\\
 & &+\,\frac{8\hbar^2(m-U/c^2)}{(m_0+m)^2(\mu_0+\mu)c^2}\frac{d{U}}{dr}\frac{\partial\Psi}{\partial{r}}\nonumber\\
 & &+\,\frac{2\hbar^2}{(m_0+m)^2(\mu_0+\mu)c^2}\left[\nabla^2\left(2mU-\frac{U^2}{c^2}\right)\right]\Psi\nonumber\\
 & &-\,\frac{1}{(m_0+m)^2(\mu_0+\mu)c^2}\left(2mU-\frac{U^2}{c^2}\right)^2\Psi\nonumber\\
 & &-\,\frac{8(m-U/c^2)}{(m_0+m)^2(\mu_0+\mu)c^2}\frac{1}{r}\frac{d{U}}{dr}\mathbf{S}\cdot\mathbf{L}\Psi+m_0c^2\Psi.
\end{eqnarray}

\begin{eqnarray}\label{j14}
  E'\psi
 &=& -\frac{4\hbar^2}{(\mu_0+\mu)(m_0+m)^2}\left(m-\frac{U}{c^2}\right)^2\nabla^2\psi\nonumber\\
 & &+\,\frac{2\mu}{(\mu_0+\mu)(m_0+m)}\left(2mU-\frac{U^2}{c^2}\right)\psi\nonumber\\
 & &+\,\frac{8\hbar^2(m-U/c^2)}{(m_0+m)^2(\mu_0+\mu)c^2}\frac{d{U}}{dr}\frac{\partial\psi}{\partial{r}}\nonumber\\
 & &+\,\frac{2\hbar^2}{(m_0+m)^2(\mu_0+\mu)c^2}\left[\nabla^2\left(2mU-\frac{U^2}{c^2}\right)\right]\psi\nonumber\\
 & &-\,\frac{1}{(m_0+m)^2(\mu_0+\mu)c^2}\left(2mU-\frac{U^2}{c^2}\right)^2\psi\nonumber\\
 & &-\,\frac{8(m-U/c^2)}{(m_0+m)^2(\mu_0+\mu)c^2}\frac{1}{r}\frac{d{U}}{dr}\mathbf{S}\cdot\mathbf{L}\psi.
\end{eqnarray}
\begin{CJK*}{GBK}{song}
(\ref{j14}) is clearly correct according to (\ref{16}). As for $\mathbf{S}\neq0$, the correctness of (\ref{j14}) still needs to be further verified.

\section{Relativistic Energy Levels for Two-Particle Systems with Zero Total Spin Angular Momentum}

A hydrogen-like atom that the total spin angular momentum is zero, and the pionium composed by $\pi^-$ and $\pi^+$, are both two-particle systems.
This type of potential energy of interaction between particles is
$U=-Ze_s^2/r,\;e_s=e(4\pi\varepsilon_0)^{-1/2}$, considering
\[\nabla^2\frac{1}{r}=-4\pi\delta(\mathbf{r}),\quad r=|\mathbf{r}_1-\mathbf{r}_2|,\]
the wave equation (\ref{16}) can be further expressed as
\begin{eqnarray}\label{18}
 E'\psi
 &=&-\,\frac{4\hbar^2}{(m_0+m)^2(\mu_0+\mu)}\left(m+\frac{Ze_s^2}{c^2r}\right)^2\nabla^2\psi\nonumber\\
 & &-\,\frac{2\mu}{(\mu_0+\mu)(m_0+m)}\left(\frac{2mZe_s^2}{r}+\frac{Z^2e_s^4}{c^2r^2}\right)\psi\nonumber\\
 & &+\,\frac{8\hbar^2}{(m_0+m)^2(\mu_0+\mu)c^2}\left(m+\frac{Ze_s^2}{c^2r}\right)\frac{Ze_s^2}{r^2}\frac{\partial\psi}{\partial{r}}\nonumber\\
 & &-\,\frac{1}{(m_0+m)^2(\mu_0+\mu)c^2}\left(\frac{2mZe_s^2}{r}+\frac{Z^2e_s^4}{c^2r^2}\right)^2\psi\nonumber\\
 & &-\,\frac{4\hbar^2}{(m_0+m)^2(\mu_0+\mu)c^2}\,\frac{Z^2e_s^4}{c^2r^4}\,\psi\nonumber\\
 & &+\,\frac{16\pi{m}\hbar^2Ze_s^2}{(m_0+m)^2(\mu_0+\mu)c^2}\,\delta(\mathbf{r})\psi.
\end{eqnarray}

Now let us solve the wave equation (\ref{18}) under the condition of
$\mathbf{r}>0$, when $\delta(\mathbf{r})=0$. Using the spherical
polar coordinates, the Laplace operator $\nabla^2$ is expressed as
\[\nabla^2=\frac{1}{r^2}\left[\frac{\partial}{\partial{r}}\left(r^2\frac{\partial}{\partial{r}}\right)+
\frac{1}{\sin\theta}\frac{\partial}{\partial\theta}\left(\sin\theta\frac{\partial}{\partial\theta}\right)+
\frac{1}{\sin^2\theta}\frac{\partial^2}{\partial\varphi^2}\right].\]

Supposing $\psi(r,\theta,\varphi)=R(r)Y(\theta,\varphi)$, substituting it into (\ref{18}) and considering $\delta(\mathbf{r})=0$, we have
\begin{eqnarray*}
  \frac{(m_0+m)^2(\mu_0+\mu)E'r^2}{4\hbar^2[m+Ze_s^2/(c^2r)]^2} &+& \frac{1}{R}\frac{d}{dr}\left(r^2\frac{d{R}}{dr}\right)
  -\frac{2Ze_s^2}{[m+Ze_s^2/(c^2r)]c^2}\frac{1}{R}\frac{d{R}}{dr}\\
  & &+\,\frac{\mu(m_0+m)r^2}{2\hbar^2[m+Ze_s^2/(c^2r)]^2}\left(\frac{2mZe_s^2}{r}+\frac{Z^2e_s^4}{c^2r^2}\right)\\
  & &+\,\frac{r^2}{4\hbar^2c^2[m+Ze_s^2/(c^2r)]^2}\left(\frac{2mZe_s^2}{r}+\frac{Z^2e_s^4}{c^2r^2}\right)^2\\
  & &+\,\frac{1}{[m+Ze_s^2/(c^2r)]^2}\frac{Z^2e_s^4}{c^4r^2}\\
    &=& -\frac{1}{Y}\left[\frac{1}{\sin\theta}\frac{\partial}{\partial\theta}\left(\sin\theta\frac{\partial{Y}}{\partial\theta}\right)
    +\frac{1}{\sin^2\theta}\frac{\partial^2Y}{\partial\varphi^2}\right]=\lambda.
\end{eqnarray*}

\begin{eqnarray}\label{19}
  & &\left(1+\frac{Ze_s^2}{mc^2r}\right)^2\frac{1}{r^2}\frac{d}{dr}\left(r^2\frac{d{R}}{dr}\right)-
  \left(1+\frac{Ze_s^2}{mc^2r}\right)\frac{2Ze_s^2}{mc^2r^2}\frac{d{R}}{dr}\nonumber\\
  & &+\,\frac{\mu(m_0+m)Ze_s^2}{m\hbar^2r}\left(1+\frac{Ze_s^2}{2mc^2r}\right)R+\frac{(m_0+m)^2(\mu_0+\mu)E'}{4m^2\hbar^2}R\nonumber\\
  & &+\left(1+\frac{Ze_s^2}{2{m}c^2r}\right)^2\frac{Z^2e_s^4}{\hbar^2c^2r^2}R
  +\frac{Z^2e_s^4}{m^2c^4r^4}R-\left(1+\frac{Ze_s^2}{mc^2r}\right)^2\frac{\lambda}{r^2}R=0.
\end{eqnarray}
\begin{equation}\label{20}
\frac{1}{\sin\theta}\frac{\partial}{\partial\theta}\left(\sin\theta\frac{\partial{Y}}{\partial\theta}\right)
+\frac{1}{\sin^2\theta}\frac{\partial^2Y}{\partial\varphi^2}+\lambda{Y}=0.
\end{equation}
According to (\ref{20}), denoting $\lambda=l(l+1),\;l=0,1,2,\ldots$,
obviously the solution of the equation is the spherical harmonics
$Y_{lm}(\theta,\varphi)$

Now let us solve the radial equation (\ref{19}), discussing the
situation of the bound state ($E'<0$). Let
\begin{equation}\label{a5}
    \alpha'=\frac{m_0+m}{m\hbar}[(\mu_0+\mu)|E'|]^{1/2},\quad\rho=\alpha'r,
\end{equation}
\begin{equation}\label{a5'}
 \beta=\frac{\mu(m_0+m)Ze_s^2}{\alpha'm\hbar^2}=\frac{Ze_s^2}{\hbar}\left[\frac{\mu^2}{(\mu_0+\mu)|E'|}\right]^{1/2},
\end{equation}
using the variable substitution $\rho=\alpha'r$, then (\ref{19}) can be expressed as:
\begin{eqnarray}\label{21}
  & &\left(1+\frac{d_0}{\beta\rho}\right)^2\frac{1}{\rho^2}\frac{d}{d\rho}\left(\rho^2\frac{d{R}}{d\rho}\right)
  -\left(1+\frac{d_0}{\beta\rho}\right)\frac{2d_0}{\beta\rho^2}\frac{d{R}}{d\rho}+\frac{\beta}{\rho}\left(1+\frac{d_0}{2\beta\rho}\right)R\nonumber\\
  &&-\,\frac{1}{4}R+\left(1+\frac{d_0}{2\beta\rho}\right)^2\frac{Z^2\alpha^2}{\rho^2}R+\frac{d_0^2}{\beta^2\rho^4}R-\left(1+\frac{d_0}{\beta\rho}\right)^2
  \frac{l(l+1)}{\rho^2}R=0.
\end{eqnarray}
Where $\alpha$ denotes the fine structure constant. $d_0$, which is a small parameter, denotes
\begin{equation}\label{d0}
  d_0=2Z^2\alpha^2D,\quad{D}=\frac{\mu(m_0+m)}{2m^2},\quad\alpha=\frac{e_s^2}{\hbar{c}}.
\end{equation}

Let $R(\rho)=u(\rho)/\rho$, considering
\[\frac{1}{\rho^2}\frac{d}{d\rho}\left(\rho^2\frac{dR}{d\rho}\right)=\frac{1}{\rho}\frac{d^2}{d\rho^2}(\rho{R}),\]
then (\ref{21}) can be expressed as:
\begin{eqnarray}\label{a6}
  & &\left(1+\frac{d_0}{\beta\rho}\right)^2\frac{d^2u}{d\rho^2}-\left(1+\frac{d_0}{\beta\rho}\right)\frac{2d_0}{\beta\rho^2}\frac{du}{d\rho}
  +\frac{2d_0}{\beta\rho^3}u+\frac{\beta}{\rho}\left(1+\frac{d_0}{2\beta\rho}\right)u\nonumber\\
  & &-\,\frac{1}{4}u+\left(1+\frac{d_0}{2\beta\rho}\right)^2\frac{Z^2\alpha^2}{\rho^2}u+\frac{3d_0^2}{\beta^2\rho^4}u
  -\left(1+\frac{d_0}{\beta\rho}\right)^2\frac{l(l+1)}{\rho^2}u=0.
\end{eqnarray}

Firstly, let us study the asymptotic behavior of this equation,
when $\rho\rightarrow\infty$, the equation can be transformed into
the following form:
\[\frac{d^2u}{d\rho^2}-\frac{1}{4}u=0,\quad{u}(\rho)=\exp(\pm\rho/2).\]
As $\exp(\rho/2)$ is in conflict with the finite conditions of wave
functions, we substitute $u(\rho)=\exp(-\rho/2)f(\rho)$ into the
equation, then we have the equation satisfied by $f(\rho)$:
\begin{eqnarray}\label{a7}
  & &\left(1+\frac{d_0}{\beta\rho}\right)^2\frac{d^2f}{d\rho^2}-\left(1+\frac{d_0}{\beta\rho}\right)
  \left(1+\frac{d_0}{\beta\rho}+\frac{2d_0}{\beta\rho^2}\right)\frac{d\,f}{d\rho}\nonumber\\
  & &+\left(\frac{\beta}{\rho}+\frac{d_0}{2\beta\rho}\right)\left(1+\frac{d_0}{2\beta\rho}\right)f
  +\left(1+\frac{2}{\rho}+\frac{d_0}{\beta\rho}+\frac{3d_0}{\beta\rho^2}\right)\frac{d_0}{\beta\rho^2}f\nonumber\\
  & &+\,\left(1+\frac{d_0}{2\beta\rho}\right)^2\frac{Z^2\alpha^2}{\rho^2}f-\left(1+\frac{d_0}{\beta\rho}\right)^2\frac{l(l+1)}{\rho^2}f=0.
\end{eqnarray}
Thus solving for the radial wave function $R(\rho)$ comes down to solving for $f(\rho)$, namely
\begin{equation}\label{22}
  R(\rho)=\frac{1}{\rho}\exp\left(-\frac{1}{2}\,\rho\right)f(\rho),\quad\rho=\frac{2Z}{\beta{a}_0}\,r,\;\;a_0=\frac{2m}{m_0+m}\frac{\hbar^2}{\mu{e}_s^2}.
\end{equation}

According to (\ref{a5'}) we have
\[\beta^2=\frac{Z^2e_s^4}{\hbar^2}\frac{\mu^2}{(\mu_0+\mu)|E'|}.\]
Substituting $\mu=\mu_0-|E'|/c^2$ into the equation above, we have
\begin{equation}\label{23}
  (Z^2\alpha^2+\beta^2)|E'|^2-2\mu_0c^2(Z^2\alpha^2+\beta^2)|E'|+\mu_0^2c^4Z^2\alpha^2=0.
\end{equation}
Solving (\ref{23}), we obtain two roots of $|E'|$:
\[|E'|=\mu_0c^2\mp\mu_0c^2\left(1+\frac{Z^2\alpha^2}{\beta^2}\right)^{-1/2}.\]
Thus we obtain the system mass $\mu$ corresponding to the reduced mass $\mu_0$:
\begin{equation}\label{23'}
    \mu=\pm\mu_0\left(1+\frac{Z^2\alpha^2}{\beta^2}\right)^{-1/2}.
\end{equation}
According to Definition 6, $\mu_0,\,\mu$ respectively denote
\[\mu_0=\frac{2m_{01}m_{02}}{m_0+m},\quad\mu=\mu_0-\frac{1}{c^2}|E'|.\]
Considering $m=m_0-|E'|/c^2,\;m_0=m_{01}+m_{02}$, $|E'|$ is derived
by (\ref{23'}). As the total energy of the system
$E=m_0c^2-|E'|=mc^2$, we can further obtain the total energy $E$ and
the system mass $m$. According to (\ref{23'}), $\mu$ has positive
and negative values. When $\mu$ takes on positive values, $E$ is
expressed as
\[E=\pm\left[m_{01}^2+2m_{01}m_{02}\left(1+\frac{Z^2\alpha^2}{\beta^2}\right)^{-1/2}+m_{02}^2\right]^{1/2}c^2.\]
When $\mu$ takes on negative values, $E$ is expressed as
\[E=\pm\left[m_{01}^2-2m_{01}m_{02}\left(1+\frac{Z^2\alpha^2}{\beta^2}\right)^{-1/2}+m_{02}^2\right]^{1/2}c^2.\]
Therefore we have two positive and two negative energy solutions.
Negative energy solutions are related to the ubiquity of antimatter,
which will not be discussed in this paper. Taking on positive energy
solutions, we have
\begin{equation}\label{24}
   E=\left[m_{01}^2+\frac{2m_{01}m_{02}}{\sqrt{1+Z^2\alpha^2/\beta^2}}+m_{02}^2\right]^{1/2}c^2.
\end{equation}
\begin{equation}\label{25}
   E=\left[m_{01}^2-\frac{2m_{01}m_{02}}{\sqrt{1+Z^2\alpha^2/\beta^2}}+m_{02}^2\right]^{1/2}c^2.
\end{equation}

Corresponding to the two positive energy solutions, the system mass $m$ has two expressions:
\begin{equation}\label{24'}
   m=\left[m_{01}^2+\frac{2m_{01}m_{02}}{\sqrt{1+Z^2\alpha^2/\beta^2}}+m_{02}^2\right]^{1/2}.
\end{equation}
\begin{equation}\label{25'}
   m=\left[m_{01}^2-\frac{2m_{01}m_{02}}{\sqrt{1+Z^2\alpha^2/\beta^2}}+m_{02}^2\right]^{1/2}.
\end{equation}
Clearly, when (\ref{23'}) takes on positive values, the system mass
$m$ is expressed by (\ref{24'}). But when (\ref{23'}) takes on
negative values, $m$ is expressed by (\ref{25'}).

In (\ref{24}) and (\ref{25}), the quantization of $E$ is mirrored by
the fact that $\beta$ is related to both the principal quantum
number $n$ and the angular quantum number $l$, i.e.
$\beta=\beta(n,l)$. Solving the equation (\ref{a7}) we obtain the
expression of $\beta(n,l)$. Therefore (\ref{24}) and (\ref{25}) are
the general expressions of the relativistic energy levels for
two-particle systems.

It seems difficult to accurately solve (\ref{a7}). Let us solve this
equation for approximate solutions to obtain the approximate
expression of $\beta(n,l)$. First, (\ref{a7}) is expressed by the
standard form of second-order ordinary differential equations
\[f''+p(\rho)f'+q(\rho)f=0.\]
Where $p(\rho),\,q(\rho)$ respectively denote
\[p(\rho)=-\left(1+\frac{d_0}{\beta\rho}\right)^{-1}\left(1+\frac{d_0}{\beta\rho}+\frac{2d_0}{\beta\rho^2}\right),\]
\parbox{16cm}{\begin{eqnarray*}
 q(\rho)
 &=& \left(1+\frac{d_0}{\beta\rho}\right)^{-2}\left[\left(\beta+\frac{d_0}{2\beta}\right)\frac{1}{\rho}+
 \left(Z^2\alpha^2-l(l+1)+\frac{d_0}{2}+\frac{d_0}{\beta}+\frac{d_0^2}{4\beta^2}\right)\frac{1}{\rho^2}\right]\\
 & &+\left(1+\frac{d_0}{\beta\rho}\right)^{-2}\left(Z^2\alpha^2-2l(l+1)+2+\frac{d_0}{\beta}\right)\frac{d_0}{\beta}\frac{1}{\rho^3}\\
 & &+\left(1+\frac{d_0}{\beta\rho}\right)^{-2}(Z^2\alpha^2-4l(l+1)+12)\frac{d_0^2}{4\beta^2}\frac{1}{\rho^4}.
\end{eqnarray*}}
Considering $d_0$ is very small, we have $p(\rho)\approx-1$, and $q(\rho)$ is approximately expressed as
\[ q(\rho)\approx\left(\beta+\frac{d_0}{2\beta}\right)\frac{1}{\rho}+
 \left(Z^2\alpha^2-l(l+1)-\frac{3d_0}{2}+\frac{d_0}{\beta}\right)\frac{1}{\rho^2}.\]
(\ref{a7}) is approximately expressed as
\begin{equation}\label{26}
  \frac{d^2f}{d\rho^2}-\frac{d\,f}{d\rho}+\left[\left(\beta+\frac{d_0}{2\beta}\right)\frac{1}{\rho}+
 \left(Z^2\alpha^2-l(l+1)-\frac{3d_0}{2}+\frac{d_0}{\beta}\right)\frac{1}{\rho^2}\right]f=0.
\end{equation}
Thus $\rho=0$ is a regular singular point of the equation (\ref{26}). Suppose the series solution of this equation can be expressed as
\begin{equation}\label{a8}
    f(\rho)=\sum^\infty_{\nu=0}b_\nu\rho^{s+\nu},\quad{b_0}\neq0.
\end{equation}
In order to guarantee the finiteness of $R=u/\rho$ at $\rho=0$, $s$
should be no less than 1. By substituting (\ref{a8}) into
(\ref{26}), as the coefficient of $\rho^{s+\nu-1}$ is equal to zero,
we have the relation satisfied by $b_\nu$:
\begin{equation}\label{a9}
    b_{\nu+1}=\frac{s+\nu-[\beta+d_0/(2\beta)]}{(s+\nu)(s+\nu+1)-l(l+1)+Z^2\alpha^2-3d_0/2+d_0/\beta}b_\nu.
\end{equation}
If the series are infinite series, then when $\nu\rightarrow\infty$
we have $b_{\nu+1}/b_\nu\rightarrow1/\nu$. Therefore, when
$\rho\rightarrow\infty$, the behaviour of the series is the same as
that of $e^\rho$, thus $f(\rho)$ in (\ref{22}) tends to infinity
when $\rho\rightarrow\infty$, which is in conflict with the finite
conditions of wave functions. Therefore, the series should only have
finite terms. Let $b_{n_r}\rho^{s+n_r}$ be the highest-order term,
then $b_{n_r+1}=0$. By substituting $\nu=n_r$ into (\ref{a9}) we
have $\beta+d_0/(2\beta)=n_r+s$. On the other hand, the series
starts from $\nu=0$, therefore, $b_{-1}=0$. Substituting $\nu=-1$
into (\ref{a9}), considering $b_0\neq0$, we have
$s(s-1)=l(l+1)-Z^2\alpha^2+3d_0/2-d_0/\beta$. Denoting $n=n_r+l+1$,
then the following set of equations can be solved for $s$ and
$\beta$:
\begin{equation}\label{a10}
    \left\{\begin{array}{l}
    s(s-1)=l(l+1)-Z^2\alpha^2+3d_0/2-d_0/\beta\\
    \beta+d_0/(2\beta)=n_r+s\\
    n=n_r+l+1
    \end{array}\right.
\end{equation}
We derive $s=1/2\pm\sqrt{(l+1/2)^2-Z^2\alpha^2+3d_0/2-d_0/\beta}$.
Considering $s$ should not be less than $1$,
$s=1/2+\sqrt{(l+1/2)^2-Z^2\alpha^2+3d_0/2-d_0/\beta}$. Thus we
obtain a specific expression of $\beta(n,l)$:
\begin{equation}\label{27}
   \beta=n-l-\frac{1}{2}-\frac{d_0}{2\beta}+\sqrt{\left(l+\frac{1}{2}\right)^2-Z^2\alpha^2+\frac{3d_0}{2}-\frac{d_0}{\beta}}=n-\sigma_l.
\end{equation}
Where $\sigma_l=l+1/2+d_0/(2\beta)-\sqrt{(l+1/2)^2-Z^2\alpha^2+3d_0/2-d_0/\beta}$.

Therefore, the relativistic energy levels for two-particle systems
(\ref{24})-(\ref{25}), the system mass (\ref{24'})-(\ref{25'}) and
(\ref{23'}) can be respectively expressed as
\begin{equation}\label{24n}
   E_n=\left[m_{01}^2+2m_{01}m_{02}\left(1+\frac{Z^2\alpha^2}{(n-\sigma_l)^2}\right)^{-1/2}+m_{02}^2\right]^{1/2}c^2.
\end{equation}
\begin{equation}\label{25n}
   E_n=\left[m_{01}^2-2m_{01}m_{02}\left(1+\frac{Z^2\alpha^2}{(n-\sigma_l)^2}\right)^{-1/2}+m_{02}^2\right]^{1/2}c^2.
\end{equation}
\begin{equation}\label{24'n}
   m=\left[m_{01}^2+2m_{01}m_{02}\left(1+\frac{Z^2\alpha^2}{(n-\sigma_l)^2}\right)^{-1/2}+m_{02}^2\right]^{1/2}.
\end{equation}
\begin{equation}\label{25'n}
   m=\left[m_{01}^2-2m_{01}m_{02}\left(1+\frac{Z^2\alpha^2}{(n-\sigma_l)^2}\right)^{-1/2}+m_{02}^2\right]^{1/2}.
\end{equation}
\begin{equation}\label{23'n}
    \mu=\pm\mu_0\left(1+\frac{Z^2\alpha^2}{(n-\sigma_l)^2}\right)^{-1/2},\quad\mu_0=\frac{2m_{01}m_{02}}{m_0+m}.
\end{equation}
\begin{equation}\label{24n'}
    \sigma_l=l+\frac{1}{2}+\frac{d_0}{2(n-\sigma_l)}-\sqrt{\left(l+\frac{1}{2}\right)^2-Z^2\alpha^2+\frac{3d_0}{2}-\frac{d_0}{n-\sigma_l}}.
\end{equation}
\begin{equation}\label{d0n}
   d_0=2Z^2\alpha^2D,\quad{D}=\frac{\mu(m_0+m)}{2m^2}.
\end{equation}

Considering a two-particle system, which is a hydrogen-like atom
composed by a spin-zero nucleus (like the deuteron) and a $\pi^-$,
called a pionic hydrogen atom, $m_{01},\,m_{02}$ are the rest mass
of $\pi^-$ and nucleus ($m_{01}\ll m_{02}$), then (\ref{24n}) can be
expanded as the following fast convergent infinite series

\parbox{16cm}{\begin{eqnarray*}
 E_n
 &=& m_{02}c^2+m_{01}c^2\left(1+\frac{Z^2\alpha^2}{(n-\sigma_l)^2}\right)^{-1/2}\\
 & &+\frac{1}{2}m_{01}c^2\frac{m_{01}}{m_{02}}\,\frac{Z^2\alpha^2}{(n-\sigma_l)^2}\left(1+\frac{Z^2\alpha^2}{(n-\sigma_l)^2}\right)^{-1}\\
 & &-\frac{1}{2}m_{01}c^2\left(\frac{m_{01}}{m_{02}}\right)^2\!\frac{Z^2\alpha^2}{(n-\sigma_l)^2}
 \left(1+\frac{Z^2\alpha^2}{(n-\sigma_l)^2}\right)^{-3/2}+\cdots.
\end{eqnarray*}}
Clearly, we obtain the normal energy levels for two-particle
systems. Using (\ref{24n}), we can calculate the energy spectrum of
pionic hydrogen atoms more accurately. The energy levels expressed
by (\ref{25n}) are called the abnormal energy levels. Unlike the
normal energy levels, the abnormal energy levels decrease with
increasing the principal quantum number $n$.

According to (\ref{24n})-(\ref{d0n}), we need to use iterative
methods for calculation. As $d_0$ is very small, taking $d_0=0$ in
the expression of $\sigma_l$, we can obtain the zeroth order
approximation of $\sigma_l$ to calculate that of $m$ and $\mu$, then
substitute them into $d_0$ to obtain its zeroth order approximation.
Substituting the zeroth-order approximation of $d_0$ and $\sigma_l$
into $\sigma_l$ to calculate its first-order approximation,
repeating this process, using the first-order approximation of
$\sigma_l$ to calculate that of $m$ and $\mu$, substitute them into
$d_0$ to obtain its first-order approximation. Substituting the
first-order approximation of $d_0$ and $\sigma_l$ into $\sigma_l$ to
calculate its second-order approximation, and repeating this process
we can calculate the nth-order approximation of $\sigma_l$, further,
we can obtain the energy levels $E_n$ and reach the required
accuracy. This calculation process can also be realized by computer
programming. Note that (\ref{23'n}) means $\mu$ has both positive
and negative values. When calculating normal energy levels, $\mu>0$,
the system mass uses (\ref{24'n}). When calculating abnormal energy
levels, $\mu<0$, $m$ uses (\ref{25'n}).

Then what is the physical meaning of abnormal energy levels?
A bound state composed of a positive and a negative pion is called
the pionium. The positronium, pionium, protonium, neutronium, etc.,
are generally called the particleium. For this type of system, we
have $Z=1$ and $m_{01}=m_{02}$. Therefore, when the pionium is at
abnormal energy levels, according to (\ref{25n}) we have
\[\lim_{n\rightarrow\infty}{E_n}\rightarrow(m_{01}^2-2m_{01}m_{02}+m_{02}^2)^{1/2}c^2=(m_{01}-m_{02})c^2=0.\]
What is the physical meaning of this result? According to
$E_n=mc^2$, we have $m=0$, which means the disappearance of the
particle system and the annihilation of a pair of positive and
negative pions. In relativistic quantum mechanics, the meaning of
the vacuum state should not be restricted to a state that the energy
is zero. For any bound state composed of a particle-antiparticle
pair, if it is at abnormal energy levels, it is in the vacuum state.
Therefore, the annihilation of a pair of positive and negative pions
has two phases. The first one composes the pionium, while the second
one is its transition from normal energy level expressed by
(\ref{24n}) to abnormal energy levels expressed by (\ref{25n}). If
this process produces $\gamma$ photons, it means a pair of positive
and negative pions annihilates into photons. The reverse process is
the pair production of positive and negative pions. A reasonable
extension of this concept is that after the annihilation of any
particle-antiparticle pair, a small percentage of the energy is
generally given to the abnormal energy levels of the particleium.
Thus this percentage of energy is also quantized, and its energy
spectrum is given by the abnormal energy levels of the particleium.
For instance, let $m_\pi$ be the rest mass of $\pi^+$, then the
abnormal energy levels of the pionium can be expressed as
\[E_n=\sqrt{2}m_\pi{c}^2\sqrt{1-\left(1+\frac{\alpha^2}{(n-\sigma_l)^2}\right)^{-1/2}}.\]
The non-relativistic approximation of this formula can be easily obtained:
\[E_n=\frac{\alpha m_\pi{c}^2}{n},\quad n=1,2\ldots\]
which is completely different from the energy spectrum of normal
matter expressed by (\ref{24n}). The abnormal energy levels expressed
by (\ref{25n}) clearly can not be applied to atoms, but may be applied to the production and annihilation of pionium.

\section{Conclusion}

In conclusion, by introducing Definition 1-6, using (\ref{1a}) and
assuming the relativistic kinetic expression is tenable on a wider
scale, the relativistic wave equations for two-particle systems is
derived, and the new relativistic two-body wave equations are
obtained. By applying this type of wave equations to pionium and
pionic hydrogen atoms, the general expression and specific
calculation formulas of relativistic energy levels for two-particle
systems are derived. Besides, we further find the relativistic
abnormal energy levels, thus the pair production and annihilation of
particles and antiparticles boil down to the transition between
normal and abnormal energy levels of two-particle systems.

\textbf{Biographies:} BI Guang-qing (1958-), male, native of Mizhi, Shaanxi, a senior teacher (Yan'an Second School, Yan'an, 716000, China), engages in abstract operators and partial diferential equations.

\end{CJK*}
\end{document}